\documentclass[showpacs,amsmath,twocolumn,showkeys,jcp,longbibliography]{revtex4-1}
\usepackage{bm}         
\usepackage{ifpdf}
\usepackage{amssymb,lineno,amsfonts}
\usepackage{graphicx}   
\usepackage{bbm}
\usepackage{mathrsfs}
\usepackage{upgreek}
\usepackage{mathtools}
\usepackage{epstopdf}
\usepackage{setspace}
\usepackage{hyperref}
\usepackage[matrix,frame,arrow]{xypic}
\usepackage{rotating}
\usepackage{float}
\usepackage{natbib}
\usepackage{color} 
\definecolor{med-blue}{RGB}{25,25,112} 
\hypersetup{colorlinks, linkcolor={red},citecolor={blue}, urlcolor={blue}}

\begin{document}
\title{Benford Analysis: A useful paradigm for spectroscopic analysis}
\author{Gaurav Bhole, Abhishek Shukla, and T. S. Mahesh}
\email{mahesh.ts@iiserpune.ac.in}
\affiliation{Department of Physics and NMR Research Center,\\
Indian Institute of Science Education and Research, Pune 411008, India}

\begin{abstract}
Benford's law is a statistical inference to predict the frequency of significant digits in naturally
occurring numerical databases. In such databases this law predicts a higher occurrence of the digit 1 in the most significant place and decreasing occurrences to other larger digits.
Although counter-intuitive at first sight, Benford's law has seen applications in a wide variety of fields like physics, earth-science, biology, finance etc. 
In this work, we have explored the use of Benford's law for various spectroscopic applications.  Although,
we use NMR signals as our databases, the methods described here may also be extended to other spectroscopic techniques. 
In particular, with the help of Benford analysis, we demonstrate the detection of weak 
NMR signals and spectral corrections.
We also explore a potential application of Benford analysis in the image-processing of MRI data.
\end{abstract}

\maketitle

\section{Introduction :}
Benford's law is an empirical law, which applies to many of the naturally occurring numerical datasets
\cite{benford}. 
It predicts the frequency of various digits in the most significant place in such datasets.
It is natural to expect that the frequency of digits 1 to 9 in the most significant place is nearly uniform,
unless the entries in the dataset are biased to one or more digits due to a constrained range of the entries.
For example, in a typical NMR spectrum normalized between 0 and 1, one might expect
a uniform distribution of digits 1 to 9 in the most significant place.
On the contrary, the Benford's law predicts a higher occurrence of the digit `1' in the most significant place, compared to other higher digits, whose occurrence decreases progressively. This non-uniform occurrence of digits in the most significant place was first observed by Simon Newcomb  \cite{newcomb} in 1881. 
However, Newcomb's article failed to gain recognition due to a lack of mathematical structure. This empirical law was rediscovered by Frank Benford \cite{benford} in 1938, who presented it with a mathematical formulation, which states that in a given dataset, the probability $P_B(d)$ of the first significant digit `$d$' is given by :
\begin{equation}
P_B(d)=\log_{10} \left(1 + \frac{1}{d} \right).
\label{eq1}
\end{equation}
This probability distribution, known as Benford's law,  predicts an occurrence as high as 30.1\%  for the digit `1' in the most significant place, whereas a mere 4.6\% for the digit 9.

\begin{figure}
\hspace*{-0.5cm}
\includegraphics[trim=0cm 5.5cm 0cm 0cm, clip=true,width=9.5cm]{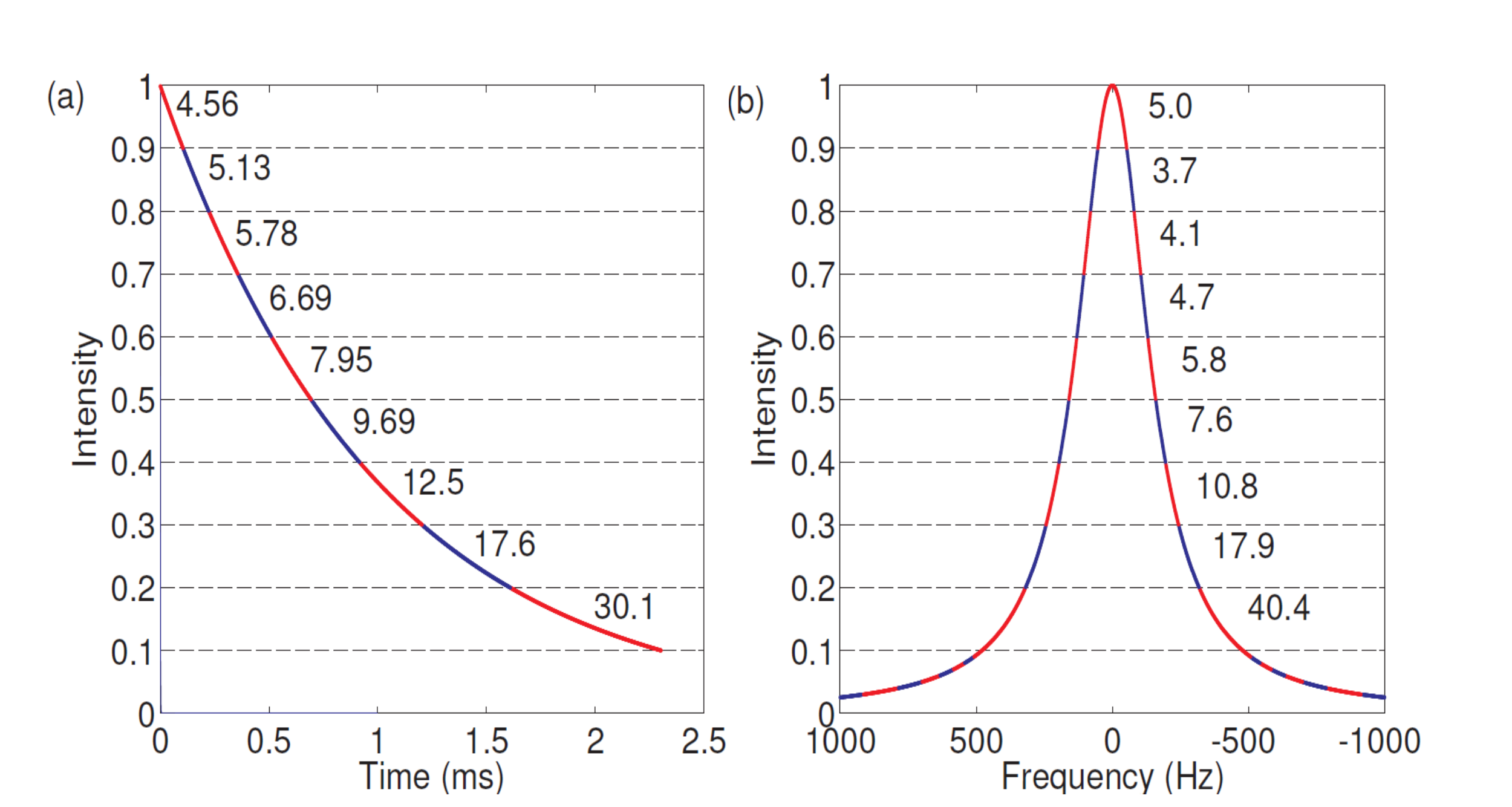} \\
\caption{(a) An exponentially decaying signal and the distribution
of various digits in the most significant place. (b) Similar distribution
for the real part of the Fourier transform of (a).}
\label{fgr:demoben}
\end{figure}

Benford's law arises from the fact that the numbering-system (be it decimal, octal, or hexadecimal) consists of mere ratios whereas many processes occurring in nature behave like a geometric progression \cite{benford}. The linear numbering-system may have 
its own advantages, but various natural phenomena such as the Maxwell-Boltzmann distribution of thermal energy, pH values of solutions, charging/discharging of capacitor, Newton's law of cooling, decay of a radioactive nuclei - all occur in exponents. 
In order to provide a simple illustration of Benford's law, we consider an exponentially decaying signal.
Such signals are versatile, including the time-domain free-induction decay (FID) of a single NMR transition.
Figure \ref{fgr:demoben}(a) displays such a digitized signal, and also marks the time-intervals at which digits 1 to 9 occur in the most significant place.  The percentage of the total occurrence of each digit is also shown
in the figure.  This distribution is in excellent agreement with the expression \ref{eq1}. Let us now consider the frequency-domain
signal which is frequently used in spectroscopic analysis.  Figure \ref{fgr:demoben}(b) displays the real part of the Fourier transform of the time-domain signal shown in Figure \ref{fgr:demoben}(a).
Such an absorptive Lorentzian line-shape is commonly found in high-resolution NMR spectroscopy.  We again notice
an overall agreement of the distribution of most significant digits with the Benford's law.

Such datasets following Benford's law also emerge from a plethora of sources ranging from astrophysical \cite{Astrophysical1}, geographical \cite{geographical}, biological \cite{neural,biology2,aerobiological}, seismographic \cite{seismic}, to financial topics \cite{stockmarket,financial1}. In the past, inspecting violations of Benford's law have proven to be useful in detecting the cases of tax fraud \cite{taxfruad} and election fraud \cite{electionfraud}.
Benford's law has also been applied to study phase transitions in quantum many-body systems \cite{ujjwal1,ujjwal2}.
Recently in another study, the possibility of distinguishing a genuine NMR spectrum from a simulated one
using Benford analysis has also been investigated \cite{gaurav}.

In this article, we explore the use of Benford's law for various spectroscopic applications, specifically in NMR. In particular, we present some useful techniques pertaining to spectroscopic peak detection, phase correction, and baseline correction. Firstly, we describe the methodology of Benford analysis. Then in the next section, 
we demonstrate three applications in NMR spectral analysis (i) the detection of weak solute peaks which are overridden by strong solvent peaks in 1D and 2D NMR, (ii) simultaneous correction of zeroth order and first order phase errors in NMR spectra, and (iii) a technique for baseline correction.  Moreover, we show that the Benford analysis of an MRI image might be helpful in highlighting its key areas.  Finally we conclude in the last section.

\section{Methodology of Benford Analysis}
We shall describe here the general procedure adopted for performing Benford analysis. Firstly, the signs of
the data entries $\{x_i\}$ are removed by taking their absolutes.  To remove a bias towards any particular digit(s)
arising from a limited data-range $[x_{\mathrm{min}},x_{\mathrm{max}}]$, the dataset is rescaled between 0 and 1,
by using the transformation \cite{ujjwal2}
\begin{eqnarray}
x_i \rightarrow \frac{x_i-x_{\mathrm{min}}}{x_{\mathrm{max}}-x_{\mathrm{min}}}.
\end{eqnarray}
The distribution $P(d)$ of digits $d = \{1,\cdots,9\}$ in the most significant place is now extracted. 
In order to quantify the amount of agreement between the observed distribution $P(d)$ with the expected distribution 
$P_B(d)$, we define `Benford Goodness Parameter' (BGP) as \cite{geographical}
\begin{eqnarray}
BGP = \left( 1 - \sqrt{ \sum_{d=1}^{9} \frac{(P(d)-P_B(d))^2}{P_B(d)}} ~\right) \times 100.
\end{eqnarray}
An ideal Benford distribution corresponds to a value of BGP = 100, whereas the datasets encountered in real life 
may take lower or even negative values.

\begin{figure}
\hspace*{-0.5cm}
\includegraphics[trim=0cm 0cm 0cm 0cm, clip=true,width=7cm]{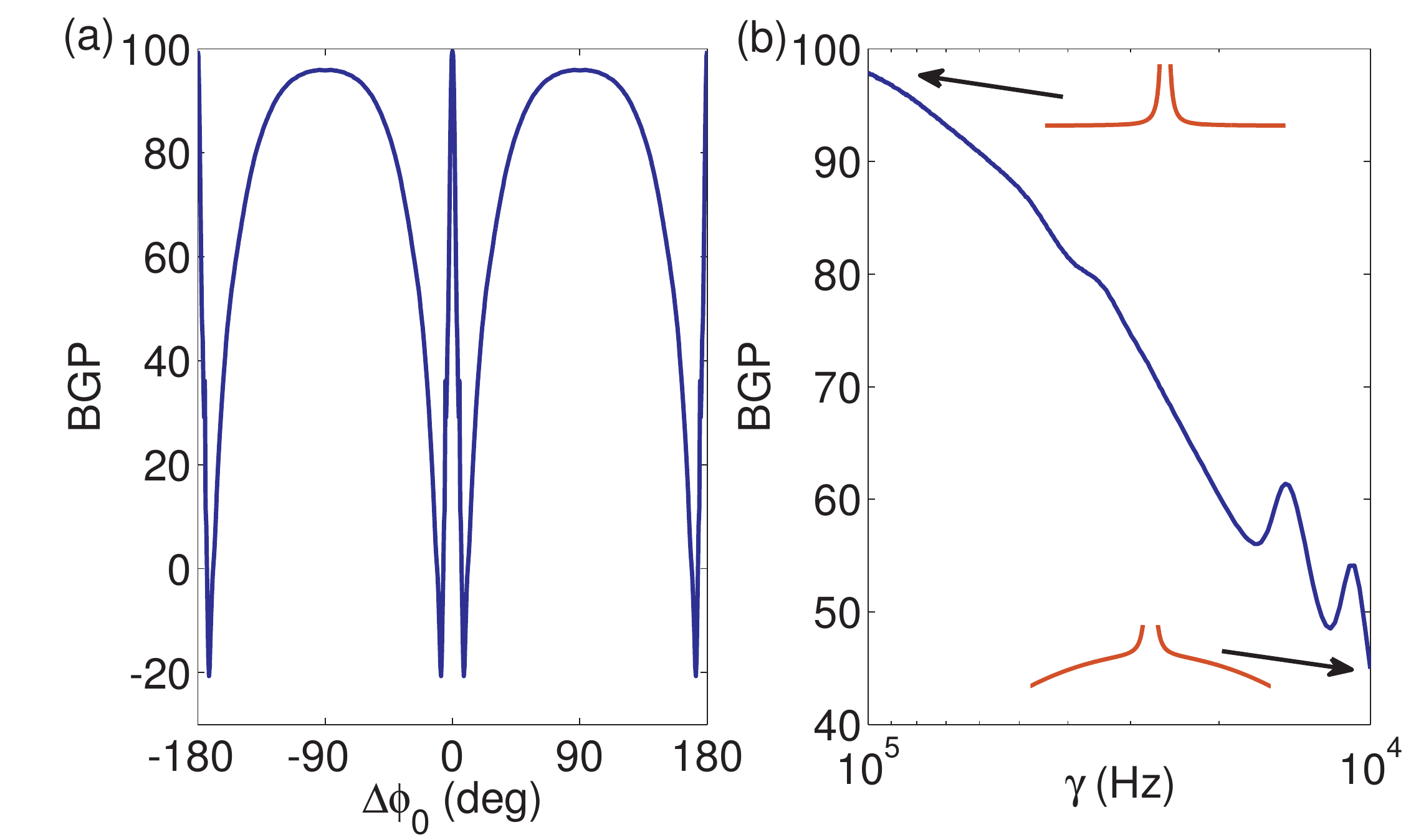} \\
\caption{BGP versus (a) phase-error and (b) the baseline-error controlled by the $\gamma$ parameter.}
\label{fgr:benbasis}
\end{figure}

Now we describe the dependency of BGP on certain spectral parameters which form the basis for
some interesting applications in spectroscopy:
(i) The BGP values depend on the spectral line-shapes and not on their relative intensities.
(ii) As illustrated in Figure \ref{fgr:benbasis}(a), the BGP value depends on the 
phase of the spectral line.  For a given Lorentzian spectral line, maximum BGP
is displayed when it is perfectly absorptive ($\phi_0 = 0$).
(iii) The BGP value also depends on the baseline.  This fact is illustrated 
in Figure \ref{fgr:benbasis}(b).  Here a Lorentzian peak is placed on a Gaussian
base line, $\exp(-\nu^2/\gamma^2)$ wherein $\nu$ is the frequency and $\gamma$ is
the width parameter. Clearly, the BGP increases as the baseline becomes flatter. 
In the following we utilize these properties of BGP to detect weak NMR peaks as well as
to rectify the phase errors and base-line artefacts in NMR spectra.

So far we have considered only one spectral line, while many spectral lines are encountered
in a typical NMR signal.  In such situations, it is helpful to consider
a `bin' formed by a smaller section of the data set and compute its BGP.  
A set of BGP values are then obtained after systematically moving the bin over 
the entire dataset.  We can then plot the BGP values versus the bin centers.  We refer
to this procedure as Scanning Benford Analysis (SBA).
The mean of the BGP values thus obtained captures the lineshapes at
all parts of the spectrum, and therefore is a measure of an overall spectral quality.
Since a typical NMR signal consists of tens of thousands of data points, 
each bin can accommodate hundreds of data points required for an
effective Benford analysis. In the case of a two-dimensional dataset, we may consider
rectangular bins and shift them systematically all over the dataset, to obtain a 2D-SBA plot.
In the following, we describe some applications of Benford analysis in NMR spectroscopy.

\section{Applications in NMR Spectroscopy:}
\subsection{Detection of weak peaks}
\begin{figure}
\includegraphics[trim=2cm 0cm 0cm 0cm, clip=true,width=5cm]{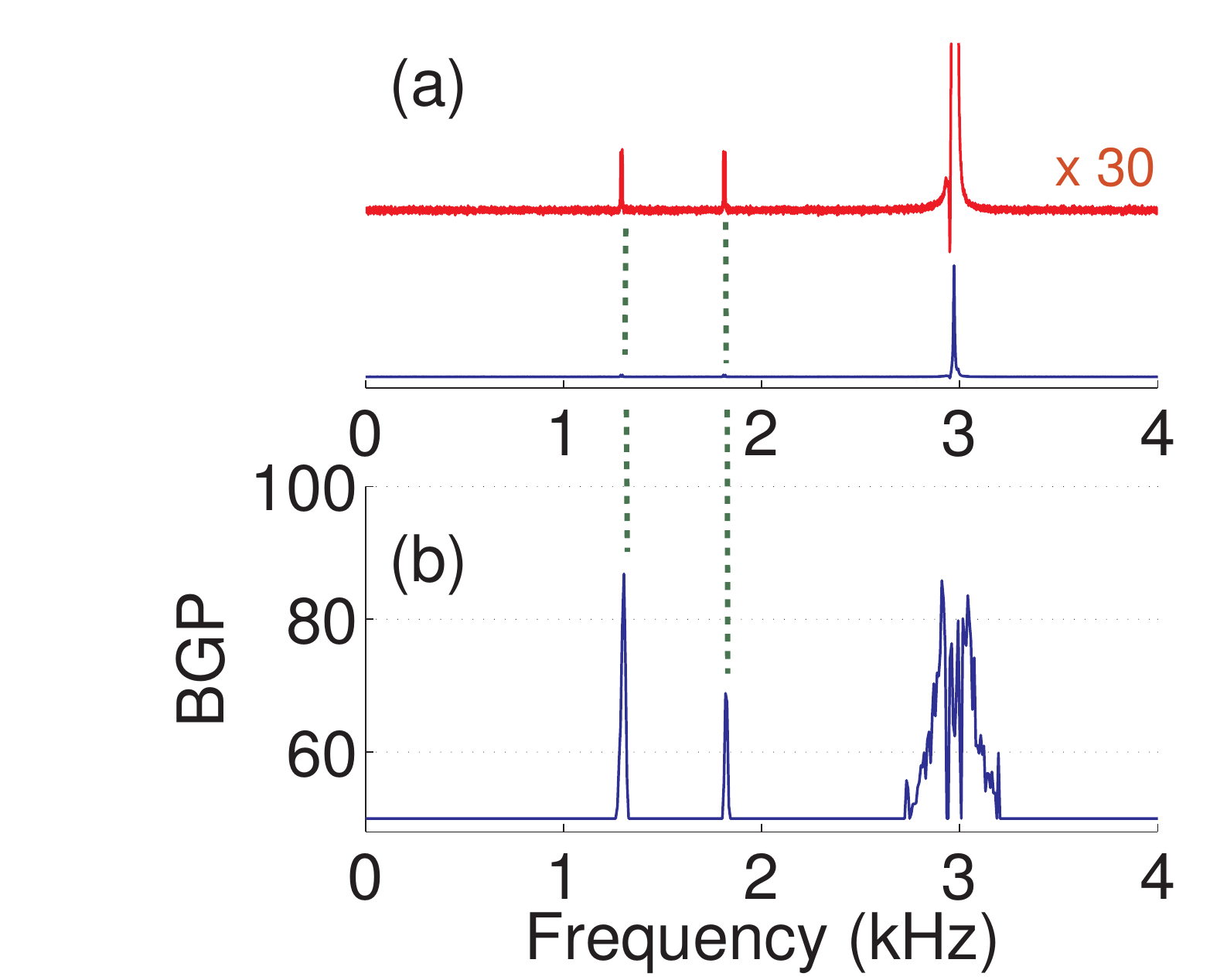} \\
\caption{The normal $^1$H spectrum of 4-Hydroxybenzoic acid in water (a) and
the corresponding SBA plot (b) (only BGP values above 50 are shown).}
\label{fgr:scan1d}
\end{figure}

As previously mentioned, BGP values are not dependent on the relative intensities of 
the spectral lines.  This is advantageous for detecting weak solute peaks overridden 
by a strong solvent peak.
To illustrate this fact, we chose a two-spin system (4-Hydroxybenzoic acid, 9.5 mg)
dissolved in D$_2$O (500 $\mu$l) and H$_2$O (50 $\mu$l).  
The $^1$H NMR signal obtained with 1 degree pulse-angle is shown in Figure \ref{fgr:scan1d}(a).  
As is evident, the strong water signal easily dominates over the solute signals.
Here the intensities of the solute peaks are about  $1.7\%$ of that of the solvent peak,
and therefore are hardly visible in the lower trace of Figure \ref{fgr:scan1d}(a). 
The SBA curve for this NMR spectrum is also shown in Figure \ref{fgr:scan1d}(b).
The SBA curve was calculated from the frequency domain signal using bins each with 1000 data points
shifted by 100 points throughout the spectrum.  For simplicity, we have truncated the SBA curve
for BGP values below 50.  Interestingly, the positions of the peaks in
the SBA plot matches fairly well with those in the NMR spectrum.  Further, the relative intensities
of the solute peaks compared to the solvent peak have also improved by about two orders of magnitude.
The fine splitting of the solvent peak in the SBA plot is mainly due to the non-Lorentzian line-shape
of the solvent peak (as can be observed in Figure \ref{fgr:scan1d}(a)).  While the resolution
of the SBA curve is limited, the sensitivity has improved significantly. 
This way, SBA curves can help in identifying weak peaks in the presence of strong peaks.

\begin{figure}
\includegraphics[trim=0cm 0cm 0cm 0cm, clip=true,width=7cm]{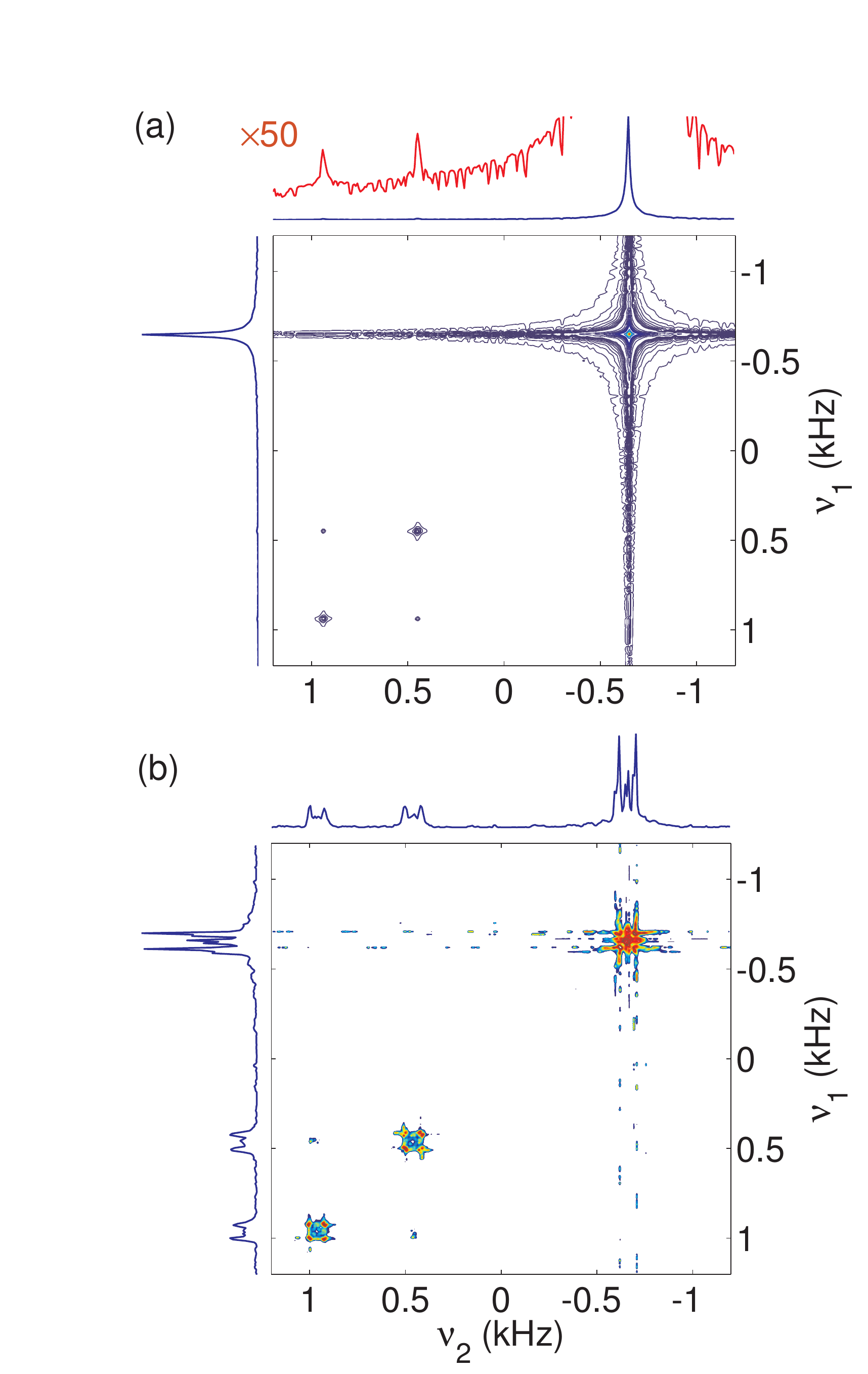} \\
\caption{(a) $^1$H 2D G-COSY spectrum of 4-Hydroxybenzoic acid in water, and
(b) the corresponding 2D-SBA contour plot.  Both (a) and (b)
are symmetrized to suppress spurious peaks. 
The horizontal and vertical projections are shown in both the cases.
}
\label{fgr:scan2d}
\end{figure}

We also illustrate the usefulness of the above procedure in multidimensional NMR
using a two-dimensional gradient COSY (G-COSY) spectrum of the same sample 
(4-Hydroxybenzoic acid).  The $^1$H-$^1$H 2D spectrum and its projections are shown
in Figure \ref{fgr:scan2d}(a). The intensities of the solute diagonal-peaks are about $1\%$ of that of the solvent peak. 
We perform a 2D-SBA of the frequency-domain spectrum by using square bins each containing $32  \times  32$ data points.
The resulting 2D-SBA plot and its projections on either axis are shown in Figure \ref{fgr:scan2d}(b). 
Again, we clearly see that the Benford analysis has given significant emphasis to the weak solute peaks such that
they appear comparable to the solvent peak and even the cross-peaks are also clearly visible in the SBA plot.
However, the fine structures in both the solvent and the solute peaks are artefacts of SBA and 
may not carry spectroscopic information.  

\begin{figure}
\hspace*{-1.2cm}
\includegraphics[trim=0cm 0cm 0cm 2.3cm, clip=true,width=9.5cm]{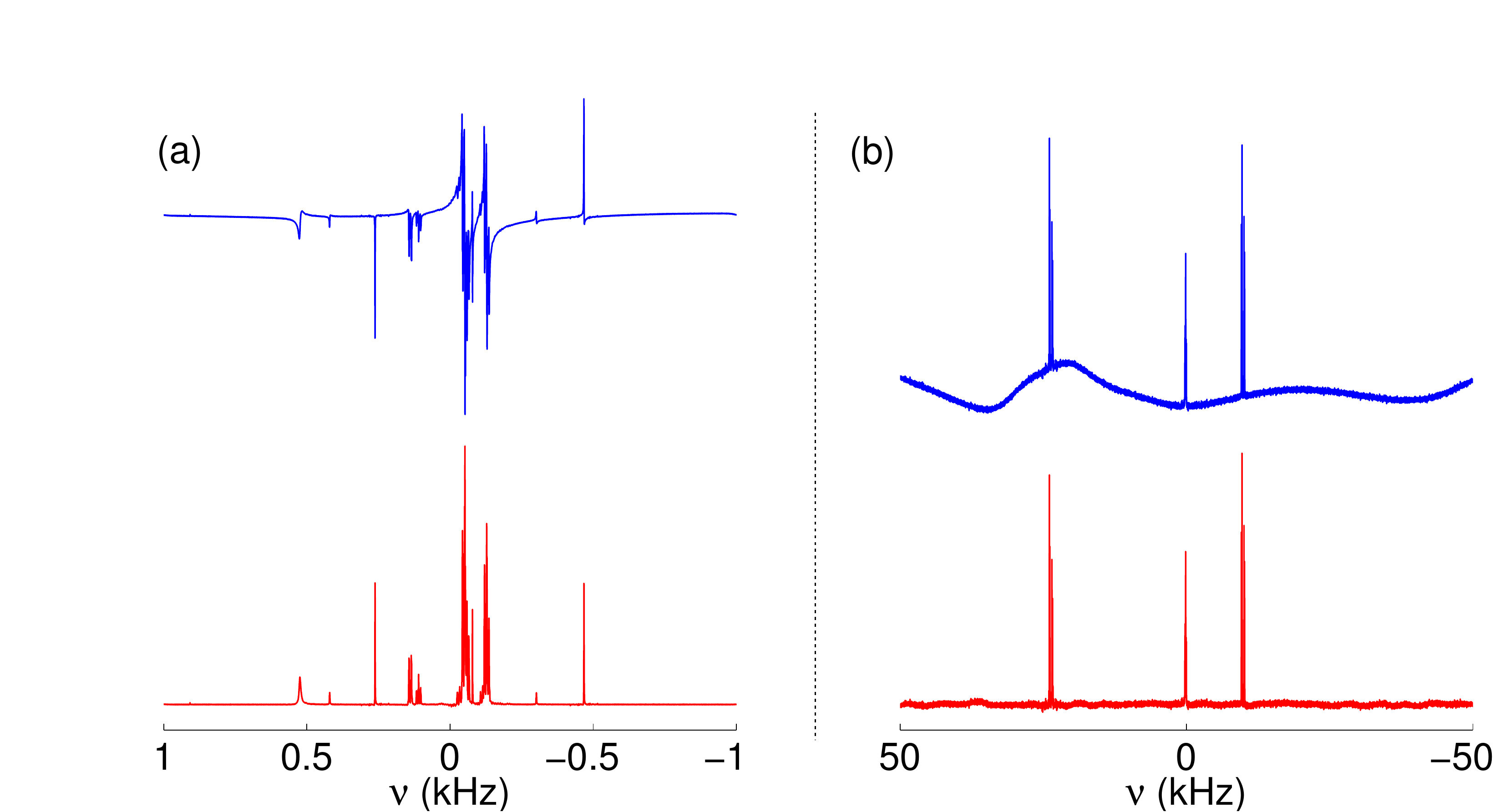} \\
\caption{Illustrating phase correction (a) and baseline correction (b) using
Benford analysis.  Figure (a) displays $^1$H spectra of Diphenoxyphosphine oxide 
(in CDCl$_3$) before (top trace) and after (bottom trace) phase correction.  Figure (b) displays
$^{19}$F spectra of 4-Bromo-1,1,2-trifluoro-1-butene (20 $\mu$l in 600 $\mu$l CDCl$_3$) before
(top trace) and after (bottom trace) baseline correction.
}
\label{fgr:phaabs}
\end{figure}

\subsection{Correction of Phase Errors}
The mismatch between the phase of the observed magnetization and the receiver phase
may lead to zeroth-order phase errors resulting in non-absorptive lineshapes.  Further,
the unavoidable delay (dead-time) introduced while switching a spectrometer from transmitter mode to
receiver mode, or certain deliberate delays after the detection pulse result in first-order phase errors.
In such cases, it is necessary
to carry out a post-processing to rectify all the phase errors. The sensitivity of  
BGP for the phase errors illustrated in Figure \ref{fgr:benbasis}(a) may be exploited for
phase correction of NMR spectrum.  The procedure involves carrying out SBA of the distorted
spectrum and computing the mean BGP over all the bins.  The phase correction is modelled in the
form $\exp(i\phi_0+\nu\phi_1)$, where $\nu$ is the frequency coordinate, and the optimum
values of zeroth-order ($\phi_0$) and first-order ($\phi_1$) corrections can be determined
by maximizing the mean BGP.  Figure \ref{fgr:phaabs}(a) illustrates the successful simultaneous correction of 
both zeroth and first order phase errors using an automated procedure described above.

\subsection{Correction of Baseline Errors}
Often the sharp high-resolution NMR spectra of solute molecules appear
mounted on broad background spectrum arising  either from solid impurities
or from the sample tube itself.  The broad humps can have arbitrary shapes and 
may interfere with quantitative spectral analysis.  In such cases, it is desirable
to suppress the broad features using post-processing.  The standard way of correcting
the baseline involves modelling the baseline using a suitable model function (polynomial or Fourier series),
and subtracting the model function from the distorted spectrum to obtain a flat baseline.
The challenging part in this method is distinguishing the genuine NMR peaks from the
baseline features.  Here, we describe a procedure for the baseline correction of an NMR 
spectrum using Benford analysis.  The method is based on the sensitivity of BGP on the 
baseline shape as illustrated in Figure \ref{fgr:benbasis}(b).
The procedure involves generating a set of spline 
model-functions (in-built in MATLAB) corresponding to different number of sampling points on the spectrum.
After subtracting these model-functions from the distorted spectrum, we obtain a set of candidate spectra.  
Performing SBA of each of the candidate spectra, and then choosing the one with maximum mean-BGP, 
we obtain the spectrum corresponding to the best baseline.  Figure \ref{fgr:phaabs}(b) illustrates
the baseline correction using an automated procedure as described above.  The main advantage
of this method over certain other methods is in omitting the requirement of manual identification
of baseline.

\begin{figure*}
\includegraphics[trim=0cm 0cm 0cm 0cm, clip=true,width=16cm]{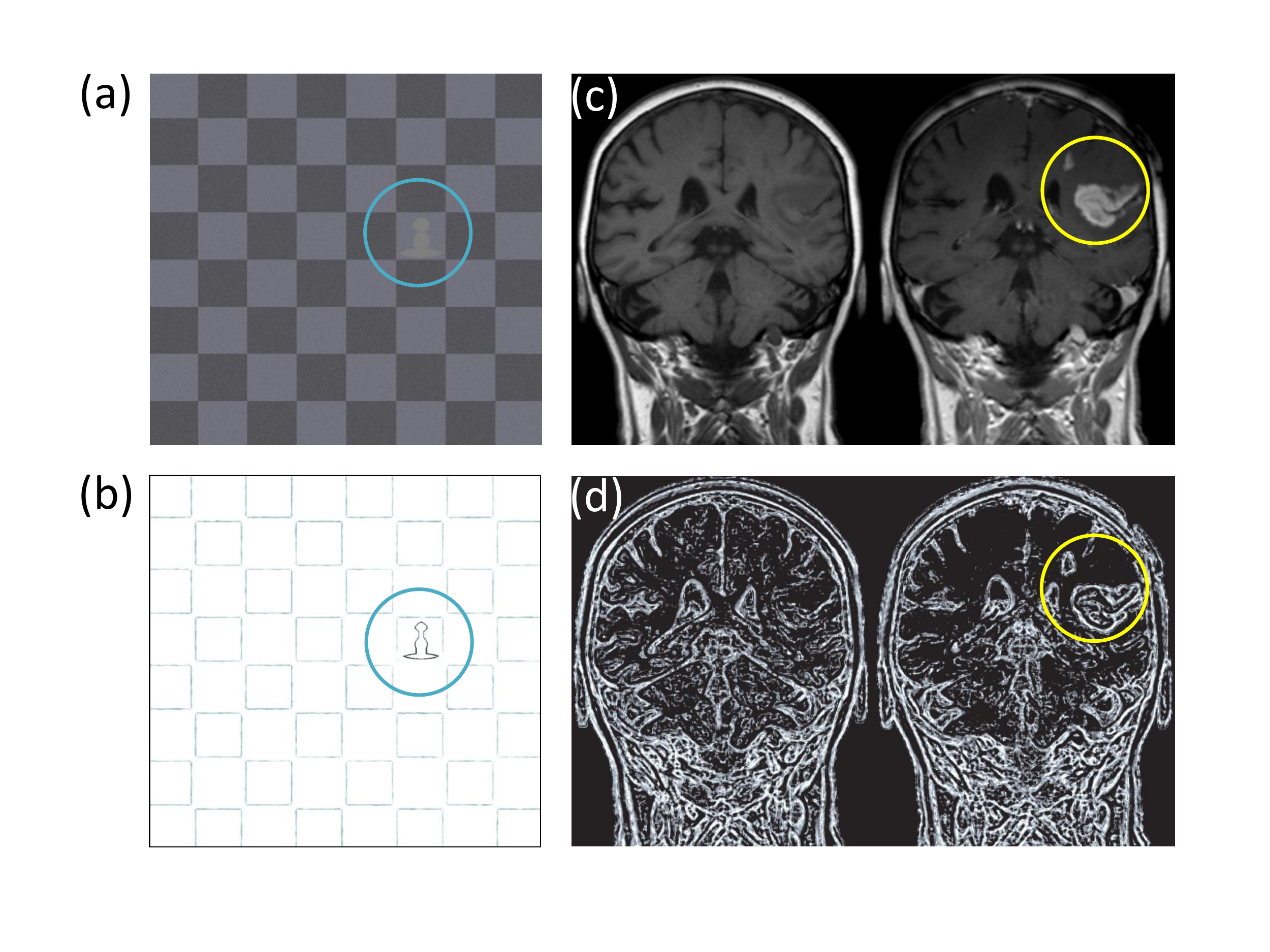} \\
\caption{(a) A chess-board with a pawn (circled), (b) 2D-SBA of (a), (c)
 2D images of MRI before (left) and after (right) introducing the contrast agent, and
(d) their corresponding 2D-SBA profiles.  The affected part is highlighted by circles.
}
\label{fgr:mri}
\end{figure*}

\section{Benford Analysis of 2D Images}
The SBA methods introduced for processing 2D-NMR spectra can also be applied to process
other 2D images. To study the effect of 2D-SBA processing, we use a noisy chess-board image 
shown in Figure \ref{fgr:mri}(a).  The corresponding 2D-SBA plot of this image, 
obtained with a bin-size of $15 \times 15$ data-points, is shown in 
Figure \ref{fgr:mri}(b).  Here we can notice two important aspects of 2D-SBA profiles.  Firstly, 
the Benford analysis clearly identifies the outlines in the image and can therefore be used 
for edge-detection.  Secondly, the weak internal features, like a faint pawn, can become
significantly prominent in the 2D-SBA plot.  This is analogous to the detection of weak NMR 
peaks in 1D-SBA and 2D-SBA described in the previous section.  These properties of 2D-SBA can
be exploited in processing many types of images including MRI.  Figure \ref{fgr:mri}(c)
displays the MRI of human brain before (left) and after (right) introducing a contrast agent
\cite{wiki}. The corresponding 2D-SBA profiles are displayed in Figure \ref{fgr:mri}(d).  
As expected, the latter images show clear outlines of all the features in (c) as well as
clearly mark-out the part highlighted by the contrast agent.  Such complementary images
may prove helpful in diagnosis.

\section{Conclusions}
The Benford's law predicts the distribution of digits 1 to 9 in the most significant
place in a numerical database.  Although it is hard to conceptualize the role of this
distribution in such a wide variety of databases, it brings out several interesting 
applications.  In this work we explored some potential applications in the analysis of 
spectroscopic data, particularly NMR data.  We found that by using a scanning Benford
analysis it is possible to detect weak peaks subdued by neighbouring strong peaks.  
We demonstrated this fact by applying the Benford analysis on both 1D and 2D NMR spectra.
The resolution of the scanning analysis is limited by the size of the bins, and
the resulting data also lacks the quantitative information on the relative intensities.
Nevertheless, such analysis may be useful for high-throughput scanning of solutes with
low-concentration.
We also demonstrated that the sensitivity of the Benford goodness parameter on 
the Lorentzian line-shape can be exploited in phase-correction as well as baseline correction.
In principle, the same property can also be used in NMR spectrometers for automated field 
homogenization.  Then we extended the Benford analysis to 2D images including an MRI image.
We found that the Benford analysis is not only capable of detecting outlines of various features in
an image but also is capable of highlighting faint portions of an image. 
We believe that the Benford analysis can open-up new vistas in spectroscopic analysis.
\section{Acknowledgements}
The authors acknowledge Ujjwal Sen and Aditi Sen, of HRI, Allahabad, 
and Anil Kumar of IISc, Bangalore, for suggesting to carry out Benford 
analysis of NMR databases.  This work was partly supported by DST project 
SR/S2/LOP-0017/2009.
\bibliographystyle{unsrt.bst}
\bibliography{ref}
\end{document}